\documentclass[fleqn,usenatbib]{mnras}

\usepackage[utf8]{inputenc}
\usepackage{enumitem}
\usepackage{graphicx}
\usepackage{hyperref}
\usepackage{mathtools}
\usepackage{comment}
\usepackage{amsmath}
\usepackage{ amssymb }
\usepackage{lipsum} 
\usepackage{float}
\usepackage{cleveref}
\usepackage[caption=false]{subfig}
\usepackage{placeins}

\title[Hubble without the Sound Horizon]{{Determining the Hubble Constant without the Sound Horizon Scale: Measurements from CMB Lensing}}

\author[E. J. Baxter \& B. D. Sherwin]{
Eric J.~Baxter$^{1,2,3,4}$
and Blake D.~Sherwin$^{3,4}$
\\$^{1}$Institute for Astronomy, University of Hawai'i, 2680 Woodlawn Drive, Honolulu, HI 96822, USA
\\$^{2}$Institute of Astronomy, University of Cambridge, Madingley Road, Cambridge CB3 0HA, UK
\\$^{3}$Department of Applied Mathematics and Theoretical Physics, University of Cambridge, Cambridge CB3 0WA, UK
\\$^{4}$Kavli Institute for Cosmology, Institute of Astronomy, University of Cambridge, Cambridge CB3 0HA, UK
}

\begin{document}

\label{firstpage}

\pagerange{\pageref{firstpage}--\pageref{lastpage}}
\maketitle

\begin{abstract}
Measurements of the Hubble constant, $H_0$, from the cosmic distance ladder are currently in tension with the value inferred from {\it Planck} observations of the CMB and other high redshift datasets if a flat $\Lambda$CDM cosmological model is assumed.  One of the few promising theoretical resolutions of this tension is to invoke new physics that changes the sound horizon scale in the early universe; this can bring CMB and BAO constraints on $H_0$ into better agreement with local measurements. In this paper, we discuss how a measurement of the Hubble constant can be made from the CMB \emph{without} using information from the sound horizon scale, $r_s$. In particular, we show how measurements of the CMB lensing power spectrum can place interesting constraints on $H_0$ when combined with measurements of either supernovae or galaxy weak lensing, which constrain the matter density parameter. The constraints arise from the sensitivity of the CMB lensing power spectrum to the horizon scale at matter-radiation equality (in projection); this scale could have a different dependence on new physics than the sound horizon. From an analysis of current CMB lensing data from {\it Planck} and \emph{Pantheon} supernovae with conservative external priors, we derive an $r_s$-independent constraint of $H_0 = 73.5\pm 5.3\,{\rm km}/{\rm s}/{\rm Mpc}$. Forecasts for future CMB surveys indicate that improving constraints beyond an error of $\sigma(H_0) = 3\,{\rm km}/{\rm s}/{\rm Mpc}$ will be difficult with CMB lensing, although applying similar methods to the galaxy power spectrum may allow for further improvements.
\vspace{2cm}
\end{abstract} 


\section{Introduction}

Measurements of the Hubble constant $H_0$, derived from observations of supernovae (SN) calibrated by the cosmic distance ladder (CDL; \citealt{Riess:2019}), are currently in tension with the value of $H_0$ inferred from {\it Planck} satellite observations of anisotropies in the cosmic microwave background (CMB; \citealt{Planck:parameters}) assuming a flat $\Lambda$CDM cosmological model.  The significance of the tension differs somewhat depending on the exact datasets considered and the statistic used, but values in the range of $4.0$--$5.8\sigma$ are typical \citep{Verde:2019}, making a statistical fluctuation an unlikely explanation \citep[e.g.,][]{Bernal:2016,Feeney:2018, Aylor:2019}.  The source of this tension is unknown, but it could conceivably result from systematic errors or error underestimates in the low-redshift (CDL) or high-redshift (CMB) measurements; an alternative explanation could be a breakdown in the cosmological model used to infer $H_0$ from the CMB.  Given these possibilities, resolving the Hubble tension is of prime importance for cosmologists.

With several groups and experiments measuring $H_0$ through different techniques, the Hubble tension has become increasingly difficult to explain by invoking systematic errors in only a single cosmological probe; arguably, it is no longer just a discrepancy between the Planck CMB and local measurements. Indeed, high values of $H_0$ close to 74 km/s/Mpc are obtained by several variants of the CDL (\citealt{Riess:2019}) as well as by strong lensing time-delay measurements (e.g., \citealt{Wong:2019}). On the other hand, a variety of probes, typically arising from earlier cosmic times, give a low value of the Hubble constant of $H_0\sim$ 67 km/s/Mpc. These include not only CMB experiments such as Planck, but also large-scale structure (LSS) probes that derive 
much of their information from the baryon acoustic oscillation (BAO) feature (e.g., \citealt{Alam:2017}). Measurements of $H_0$ from galaxy BAO can be obtained in combination with CMB datasets -- with {\it Planck}, WMAP, ACT, and SPT combined with BOSS BAO giving similar results -- or even derived independently of CMB anisotropy measurements, by combining BAO analyses with information about the baryon density inferred from big bang nucleosynthesis (BBN) measurements \citep{Addison:2018}.  The $H_0$ constraints from combining galaxy BAO and BBN can be tightened further by including external $\Omega_m$ information from e.g., a weak lensing survey \citep{Abbott:2018}, or by including Lyman-alpha BAO \citep{Addison:2018, Cuceu:2019}. In all cases, a value of the $H_0$ is obtained that is in agreement with the Planck inference (see e.g., \citealt{Aylor:2019} and \citealt{Abbott:2018} for discussion of this point).

We note that since supernovae, galaxy BAO and Lyman-alpha BAO provide measurements of the expansion history over a wide range of redshifts, it appears very difficult to explain the Hubble tension by modifying the dynamics of the universe at late times (e.g. \citealt{Feeney:2018}, \citealt{Aylor:2019}).

Interestingly, as pointed out by others \citep[e.g.][]{Bernal:2016,Aylor:2019},  all of the probes that yield a low value of $H_0$ close to 67 km/s/Mpc have one important aspect in common: they all assume that the calculation of the sound horizon scale $r_s$ from the standard cosmological model is correct ($r_s$ is the distance travelled by a sound wave until CMB last scattering).

In the case of the CMB, for instance, the calibration of $H_0$ is derived directly from the measurement of the angular scale subtended by $r_s$. CMB power spectrum measurements constrain the physical energy densities of baryons ($\Omega_b h^2$) and matter ($\Omega_m h^2$), where $h$ is the Hubble constant measured in units of $100~\mathrm{km/s/Mpc}$.  Assuming the standard cosmological model and assuming our understanding of pre-recombination physics is correct, these constraints imply a value of the sound horizon scale $r_s$. Since the CMB power spectrum also directly measures the angular scale $\theta_s$ subtended by $r_s$ via the peak spacing, one can combine the calibrated physical size and measured angular size of $r_s$ to constrain the comoving distance to the last scattering surface, $\chi_* = r_s/\theta_s$.  In flat $\Lambda$CDM cosmological models, $\chi_*$ depends only on $\Omega_m$ and $H_0$.  The CMB constraints on $\Omega_m h^2$ and on $\chi_*(\Omega_m, h)$ then together break the degeneracy between $h$ and $\Omega_m$, leading to a constraint on $h$.  Similarly, the measured angular and redshift scales of $r_s$ from the BAO feature are used to infer $H_0$ assuming a standard calibration of the physical scale of $r_s$ from CMB or BBN measurements.\footnote{To be more precise, the BAO feature is connected to the sound horizon at the end of the baryon drag epoch, not at last scattering.  However, since their dependence on cosmology is nearly identical, we neglect the small difference between these scales here and use $r_s$ to refer to the sound horizon scales at both last-scattering and at the end of the baryon drag epoch. }

Consequently, some of the most promising proposed theoretical resolutions to the Hubble tension rely on modifying $r_s$ by invoking new physics prior to last scattering.  Proposed solutions include modifications to the expansion history via early dark energy (e.g., \citealt{Poulin:2019, Agrawal:2019}; although we note that issues were recently raised with these solutions by \citealt{Hill:2020, Ivanov:2020b, dAmico:2020b, Krishnan:2020}), changes in neutrino sector physics (e.g., \citealt{Kreisch:2019}), or changes to recombination (e.g., \citealt{Chiang:2018}).

Motivated by these considerations, in this work we explore how measurements of the CMB lensing power spectrum can be used to infer $H_0$ in a way that is independent of $r_s$. Weak gravitational lensing of a light source (e.g., galaxies or the CMB) occurs when photons from the source encounter gravitational potential wells along the line of sight and experience a gravitational deflection at each well.  The weak lensing power spectrum reflects the net impact of all of these deflections.  Consequently, changing the Hubble constant impacts the weak lensing power spectrum in several ways: (1) by changing the matter power spectrum such that the size and depth of potential wells along the line of sight is impacted, (2) changing the distance to the source so that the number of deflections experienced by a photon from the source is changed, (3) changing the apparent angular size of each deflection.

As we discuss in more detail below (and as was pointed out by \citealt{Plancklensing2015}), the net result of these effects is that the CMB lensing power spectrum is sensitive to $k_{\rm eq} \chi_* \propto \Omega_m^{0.6}  h$, where $k_{\rm eq}$ is the wavenumber corresponding to the horizon size at matter-radiation equality, and $\chi_*$ is the comoving distance to the last scattering surface.  Given a very conservative prior on $A_s$, one can determine this parameter combination from the CMB lensing power spectrum.  Then, combining with an external measurement of $\Omega_m$ to break the degeneracy, we can place a constraint on $H_0$.  Since neither $k_{\rm eq}$ nor $\chi_*$ depend on the sound horizon scale, as long as the $\Omega_m$ prior is obtained independently of the sound horizon, the constraint on $H_0$ will also be $r_s$-independent.  

The measurement of $H_0$ considered here effectively uses the broadband shape of the matter power spectrum --- which depends on $k_{\rm eq}$  --- to constrain $H_0$ rather than any features related to the sound horizon scale. Since $k_{\rm eq}$ probes the Universe at somewhat earlier times than the sound horizon at CMB last-scattering, $r_s$, and since many of the most commonly considered new physics models modify the dynamics (i.e. the expansion rate) soon before recombination, it is reasonable to expect that $k_{\rm eq}$ and $r_s$ could be differently affected by new physics at the few-percent level. Disagreement of $H_0$ measurements via $k_{\rm eq}$ (e.g., with CMB lensing) and via $r_s$ could therefore indicate that new physics may be responsible for the Hubble tension, whereas agreement at high precision would be a valuable consistency test for the standard $\Lambda$CDM model.   

As a caveat, we note that the description of the physics in terms of the sound horizon and equality scales is only a simplified one.  The full calculation starting from the perturbation equations does not involve these scales; rather, they emerge from the evolution equations if several approximations are made.  We emphasize that while we motivate our methodology using these emergent scales, when performing forecasts or analyzing current data we rely on the full calculation of the evolution of perturbations in $\Lambda$CDM, computed using a Boltzmann code.   Furthermore, even though the description in terms of $r_s$ and $k_{\rm eq}$ is approximate, due to its physical motivation, our $H_0$ measurement should be differently sensitive to many tension-motivated new physics models; a comparison with the standard, $r_s$-derived measurements will therefore give a valuable consistency test of the standard model.  A complete analysis of any new physics model, however, requires implementing its effects directly in the perturbation equations.

One could also use observables besides CMB lensing --- such as galaxy  clustering --- to infer the broadband matter power spectrum and constrain $H_0$ in a similar way to that described above.  CMB lensing has the advantage that it is naturally  sensitive to a wide range of $k$ (especially low $k$), which improves constraints on $H_0$ from the broadband information in the power spectrum. Since it probes the mass distribution directly, CMB lensing also avoids the complications and parameter degeneracies associated with galaxy biasing. Furthermore, and most importantly, because lensing is a line-of-sight integrated quantity, it is naturally less sensitive to features in the power spectrum that depend on $r_s$, such as the BAO oscillations and the scale of baryonic suppression (see discussion in \S\ref{sec:parameter_dependence}).  Extracting information about $H_0$ from the galaxy power spectrum in a way that does not depend on $r_s$ is nontrivial because of these effects (see discussion in \S\ref{sec:discussion}).  

Recently, several authors have presented high-precision, CMB-independent constraints on $H_0$ derived from the full shape of the galaxy power spectrum measured by BOSS \citep[e.g.][]{Philcox:2020, Ivanov:2020a, dAmico:2020a}.  In some sense, these constraints are similar to those considered here in that they also derive a fraction of their information about $H_0$ from the apparent scale of $k_{\rm eq}$.  However, as we discuss in more detail in \S\ref{sec:otherProbes}, for the three dimensional galaxy clustering measurements analyzed by these authors, information about $H_0$ also enters via the BAO feature and potentially also via the scale of baryonic suppression in the full-shape matter power spectrum.  Both the BAO feature and the baryon suppression scale are sensitive to $r_s$ (for the BOSS galaxy power spectrum, the former is the dominant source of information), making the resultant $H_0$ constraints sensitive to $r_s$ information as well.  In this work, we explicitly ensure that no information from $r_s$ informs our $H_0$ constraints.

Our paper presents an $r_s$-independent constraint on the Hubble constant derived from the CMB lensing power spectrum measured by {\it Planck} \citep{Plancklensing2015} in conjunction with a supernovae measurement of $\Omega_m$ from \cite{Scolnic:2018} and motivated external priors.  As discussed in \S\ref{sec:current_data}, our result is a constraint of $H_0 = 73.5\pm 5.3\,{\rm km}/{\rm s}/{\rm Mpc}$. We also present forecasts for future experiments.  These forecasts serve both to illustrate important parameter degeneracies and to show how constraints will improve with future data.  

The paper is organized as follows: in \S\ref{sec:parameter_discussion} we discuss how the lensing power spectrum depends on the $\Lambda$CDM parameters, followed by a brief discussion of why new physics may differently affect the sound horizon and equality scales in \S\ref{section:newPhysics}. In \S\ref{sec:forecast_setup} we present the forecasting assumptions and discuss the datasets involved.  For pedagogical reasons, we first present in \S\ref{sec:results} the  forecast constraints on the Hubble constant for future experiments; these forecasts serve to illustrate how parameter degeneracies can be broken.  We present our $r_s$-independent constraint on the Hubble constant using current data in \S\ref{sec:current_data} and discuss prospects for improvements with other probes in \S\ref{sec:otherProbes}.  We summarize our results in \S\ref{sec:discussion}.

\section{Constraining \texorpdfstring{$H_0$}{Lg} with the CMB lensing power spectrum}
\label{sec:parameter_discussion}

We begin by discussing how the CMB lensing power spectrum responds to the parameters of flat $\Lambda$CDM cosmological models.  The lensing power spectrum can be related to the matter power spectrum via the Limber approximation \citep{Limber:1953,Loverde:2008}:
\begin{multline}
\label{eq:limber}
C^{\kappa\kappa}(L) = \left( \frac{3\Omega_m H^2_0}{2c^2} \right)^2 \\
\int d\chi \frac{1}{a(\chi)^2} \left( \frac{\chi_* - \chi}{\chi_*} \right)^2 P\left(\frac{L+1/2}{\chi}, \chi \right),
\end{multline}
where $\chi$ is the comoving distance along the line of sight, $\chi_*$ is the comoving distance to the last scattering surface, $a(\chi)$ is the scale factor, $\Omega_m$ is the matter density parameter today, $H_0$ is the Hubble constant, and $P(k,\chi)$ is the matter power spectrum as a function of wavenumber $k$ and comoving distance $\chi$.  Although prefactors of $\Omega_m$ and $H_0$ enter into Eq.~\ref{eq:limber}, this is somewhat misleading: these prefactors result from the conversion of the matter power to the lensing potential which sources the lensing deflection.  Since the matter power spectrum is conventionally defined in terms of the primordial potential power spectrum, we will see that these factors of $\Omega_m$ and $H_0$ are cancelled by corresponding factors in $P(k,z)$.

\subsection{The broadband shape of the matter power spectrum}
\label{sec:MRE}

Because the CMB lensing power spectrum is related to an integral over the matter power spectrum, the lensing power is sensitive to the broadband shape of $P(k,z)$.  The location of the peak of $P(k,z)$ as a function of $k$ is controlled by the scale of matter-radiation equality (MRE).  Small scale modes enter the horizon before MRE and only grow logarithmically during radiation domination; this different behaviour arises because the radiation perturbations experience pressure forces and oscillate on scales below their Jeans scale, which matches the particle horizon during radiation domination. This leads to the roughly $k^{-3}$ dependence of $P(k)$ at large $k$.  Large scale modes, on the other hand, enter the horizon during matter domination (when the relevant Jeans scale has become extremely small) and the perturbation amplitude grows in proportion to the scale factor on sub-horizon scales, preserving the form of the initial power spectrum.  For a scale-invariant initial spectrum, in the small $k$ limit the power spectrum goes as $P(k) \propto k$.  Consequently, the power spectrum has a break at $k_{\rm eq}$, where $k_{\rm eq}$ is the comoving wavenumber corresponding to the horizon size (or Jeans scale) at MRE.

The redshift of MRE can be related to the present-day physical matter and radiation densities, $\omega_m$ and $\omega_{r}$, respectively.  Given the different scaling of $\omega_m$ and $\omega_{r}$ with redshift, we have $1+z_{\rm eq} = \omega_m/\omega_{r}$.  Since $\omega_{r}$ is fixed by $T_{\rm CMB}$ (assuming standard model neutrino properties and thermal history), and $T_{\rm CMB}$ is tightly constrained by e.g., COBE \citep{Fixsen:1996}, the redshift of MRE can be directly related to the physical matter density, $\omega_m$.  The horizon scale at MRE is then 
\begin{equation}
d_{\rm eq} = \int_{z_{\rm eq}}^{\infty} \frac{dz}{H(z)} \propto \frac{1}{\omega_m},
\end{equation}
giving $k_{\rm eq} \propto \omega_m$.\footnote{Strictly speaking, the scale $k_{\rm eq}$ is typically defined via the Hubble radius, i.e. $k_{\rm eq} \equiv a_{\rm eq}H(a_{\rm eq})$, rather than the true particle horizon scale at MRE.  However, the physically relevant scale is the Jeans scale, which coincides with the particle horizon scale during radiation domination.  Since during radiation domination the Hubble radius and particle horizon scales are also equivalent, the difference between defining $k_{\rm eq}$ via the Hubble radius, the particle horizon scale, or the Jeans scale is of minimal importance for our purposes.}  Increasing $\omega_m$ will make MRE occur earlier, which will shift the turnover in the matter power spectrum to smaller scales.  It is also possible to express $k_{\rm eq}$ in terms of the density parameter, $\Omega_m \propto \omega_m/\rho_{\rm crit}$, which yields $k_{\rm eq} \propto \Omega_m h^2$.  Note, though, that the apparent dependence of $k_{\rm eq}$ on $h$ is somewhat misleading, as it really enters via the relation between the critical density today and $H_0$.

\subsection{A simplified model for the CMB lensing power spectrum}
\label{sec:parameter_dependence}

We now derive simple expressions for the parameter dependencies of the CMB lensing power spectra which can be used to build intuition about how various combinations of datasets can be used to constrain $H_0$.  The discussion in this section relies heavily on \citet{Pan:2014} and Appendix E of \citet{Plancklensing2015}. 

The linear power spectrum of the density contrast can be written as
\begin{equation}
\label{eq:linear_power}
P_{\rm lin}(k,z) \propto \frac{1}{(\Omega_m H_0^2)^2} A_s k^{n_s} T^2(k) a^2(z) G^2(z),
\end{equation}
where $T(k)$ is the transfer function and $G(z)$ is the linear growth factor, and the proportionality  includes numerical factors but no cosmological dependence. As alluded to earlier, the prefactors of $\Omega_m$ and $H_0$ cancel with corresponding factors in Eq.~\ref{eq:limber}.  As is clear from Eq.~\ref{eq:limber}, large $L$ in the lensing power spectrum corresponds to large $\chi$, or high redshift for the peak of the matter power spectrum.  As a result, for $L > 100$, the CMB lensing power spectrum is sensitive to sufficiently high redshift that we can take the growth factor to be $G(z) \sim 1$ \citep{Pan:2014}.

We can approximate the transfer function over some $k$ range as \citep{Pan:2014}
\begin{equation}
\label{eq:tf_approx}
T(k) \sim \left(  \frac{k}{k_{\rm eq}}\right)^{-c},
\end{equation}
where $c \sim 2$ for $k \gg k_{\rm eq}$ and $c \sim 0$ for $k \ll k_{\rm eq}$. Substituting Eq.~\ref{eq:linear_power} and Eq.~\ref{eq:tf_approx} into the Limber equation (Eq.~\ref{eq:limber}), we have
\begin{multline}
\label{eq:cmb_limber_tfapprox}
C^{\kappa\kappa}(L) \propto \\
A_s L_{\rm eq} k_{\rm eq}^{n_s-1} (L/L_{\rm eq})^{n_s-2c}  \int dx  (1-x)^2 x^{2c-n_s}, 
\end{multline}
where we have defined $L_{\rm eq} \equiv k_{\rm eq} \chi_*$ and have substituted $x =  \chi/\chi_*$.   For the concordance $\Lambda$CDM cosmological model $L_{\rm eq} \sim 140$.  It is clear from Eq.~\ref{eq:cmb_limber_tfapprox} that for $n_s \approx 1$, the main parameter dependence of the lensing power spectrum enters via $L_{\rm eq}$ and an overall normalization by $A_s$.  Furthermore, the shape of the power spectrum (i.e. the $L$-dependent part) is controlled by $L_{\rm eq}$, $n_s$ and the transfer function (via the power law index $c$). 

In $\Lambda$CDM cosmological models, $\chi_*$ depends only on $\Omega_m$ and $h$ (at fixed $z_*$).  For small variations around a fiducial choice of $\Omega_m$, one can approximate the parameter dependence of $\chi_*$ with a power law function of $\Omega_m$: $\chi_* \propto \Omega_m^{\alpha} h^{-1}$, where $\alpha \approx -0.4$ \citep{Plancklensing2015}. This gives
\begin{equation}
\label{eq:Leq}
L_{\rm eq} \equiv k_{\rm eq} \chi_* \propto \Omega_m^{\alpha+1} h \propto  \Omega_m^{0.6} h.
\end{equation}

\subsection{\texorpdfstring{$r_s$}{Lg}-independent constraints on \texorpdfstring{$H_0$}{Lg} from the CMB lensing power spectrum}
\label{sec:constraint_explanation}

In the small-scale limit ($L \gg L_{\rm eq}$) we have $c \sim 2$; additionally adopting $n_s \sim 1$, Eq.~\ref{eq:cmb_limber_tfapprox} becomes
\begin{equation}
\label{eq:cmb_limber_ns1}
C^{\kappa\kappa}(L) \propto A_s L_{\rm eq} (L/L_{\rm eq})^{-3}.
\end{equation}
To a good approximation, then, CMB lensing constrains $H_0$ via $L_{\rm eq}$, and we expect to see an $h \propto \Omega_m^{-0.6}$ degeneracy when performing a $\Lambda$CDM fit to the CMB lensing power spectrum.  We also see that the information about $L_{\rm eq}$ is degenerate with $A_s$ (at least at small scales where the above approximations apply).  We discuss the accuracy of the approximations made above and provide more physical intuition for the parameter dependence of the lensing power spectrum in Appendix~\ref{app:lensing_approx}.

Our approach to constraining $H_0$ without $r_s$ information is therefore as follows.  We use a conservative external constraint on $A_s$ in conjunction with the CMB lensing power spectrum to constrain $L_{\rm eq}$.  Next, we use an external constraint on $\Omega_m$ to break the $\Omega_m$-$H_0$ degeneracy and constrain $H_0$.  In both cases, we use external constraints that are not sensitive to $r_s$.  Since the $L_{\rm eq}$ information from the CMB lensing power spectrum is independent of sound horizon physics, and since the external constraints are as well, the final constraint on $H_0$ is obtained without relying on the sound horizon scale.  

We note that the approximations leading to Eq.~\ref{eq:cmb_limber_ns1} break down at low $L$, leading to a departure from pure power law behavior.  Consequently, the {\it shape} of the CMB lensing power spectrum at low $L$ serves to break what would otherwise be a complete degeneracy between $A_s$ and $L_{\rm eq}$ in setting the amplitude of the CMB lensing power spectrum.  For current data, for which a significant fraction of the information comes from low $L$, the degeneracy between $A_s$ and $L_{\rm eq}$ does not have a very large impact on our $H_0$ constraints.  For future measurements, however, more of the information comes from high $L$, causing the $A_s$ prior to become more important for obtaining tight $H_0$ constraints.  For both current and future data, we will find that our results are not very sensitive to the $A_s$ prior, likely because of the low-$L$ shape information.

\subsection{Additional complications}

We have until now ignored the impact of baryons on the matter power spectrum.  Baryons introduce acoustic wiggles into the power spectrum, but these are washed out by the line of sight integration in Eq.~\ref{eq:limber}, and so do not contribute substantially to the Hubble constraints.  However, an additional impact of baryons is to suppress the power spectrum at scales below the sound horizon, $r_s$  (again, we are ignoring the small difference in redshift between the end of the drag epoch and last scattering).  After MRE but before the end of the photon-baryon drag epoch, the growth of fluctuations in the baryons is suppressed relative to the CDM because the baryons are coupled to the radiation and experience pressure effects \citep[e.g.,][]{EH:1998}.  The net result is a suppression of the total matter power at scales smaller than the sound horizon at the end of the Compton drag epoch, roughly $r_s$.  We note that for reasonable values of $\Omega_m h^2$, $r_s$ is smaller than the horizon scale at MRE \citep{Eisenstein:1998}.

One might worry that information about $r_s$ could enter into an analysis of the CMB lensing power spectrum via the scale of baryonic suppression of the matter power spectrum, which could in turn could lead to an $r_s$-dependent inference of $H_0$.  For our purposes, this is not desirable since we are explicitly trying to avoid a sound horizon-dependent measurement of $H_0$.  We will take care to demonstrate that this is not occurring for our constraints.  In essence, this is because the projection inherent in the CMB lensing power spectrum smoothes over the sharp baryonic features in the matter power spectrum.\footnote{We also neglect effects from baryonic feedback on the matter power spectrum in our analysis, making the approximation that a CMB lensing analysis that focuses on deriving $k_{\rm eq}$ should be fairly insensitive to small-scale feedback effects for the purposes of forecasting; however, a full analysis with future experiments should account for baryonic feedback, for example, by marginalizing over appropriate templates.}

The above discussion has also ignored the impact of nonlinear evolution on the matter power spectrum, as well as the impact of massive neutrinos.   Because the CMB lensing power spectrum mainly probes large scales and the high redshift Universe, it is mostly sensitive to mildly nonlinear physics for current and next generation surveys.  Below, we will include nonlinear contributions to the matter power spectrum in our forecasts using \texttt{HALOFIT} \citep{Smith:2003} with the \citet{Bird:2012} extension to include the impact of massive neutrinos.  Neutrinos do not cluster at small scales, leading to a scale-dependent suppression of the matter power spectrum (see e.g. \citealt{Lesgourgues:2006} for a review).  We expect this suppression to be somewhat degenerate with the effects of other parameters and to therefore degrade our constraints on $H_0$ somewhat.  As we discuss in \S\ref{sec:forecast_setup}, we marginalize over a free neutrino mass in our analysis.

Finally, we note that some knowledge of the CMB power spectrum is required to correctly normalize the lensing power spectrum. We make the approximation that we may neglect any changes in the normalization of the lensing power spectrum due to changes in the CMB power spectrum.  This is only a very weak requirement, which is almost trivially satisfied: since the precision of the CMB lensing power spectrum is much lower than that of the CMB power spectum, this requires only that any new physics considered must not produce a CMB power spectrum that is grossly inconsistent (i.e., differing by an amount comparable to the lensing power spectrum fractional error) with current observations.

\begin{figure}
    \centering
    \includegraphics[width=\columnwidth]{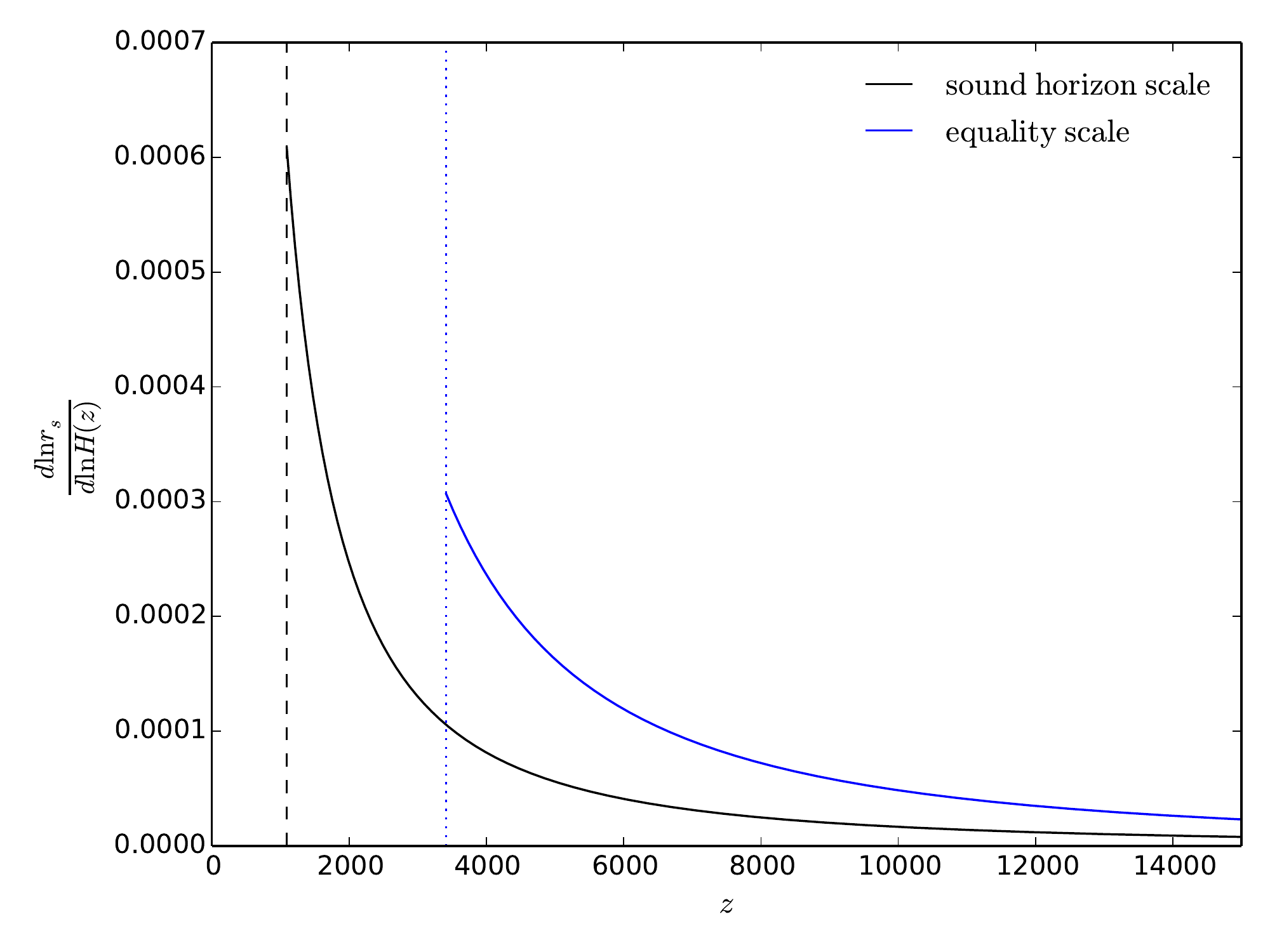}
    \caption{Illustration of the effect of changes to the Hubble rate at different redshifts on the sound horizon scale at CMB last scattering (black solid line at left) and the matter-radiation equality scale (blue solid line at right). Dashed and dotted lines indicate the redshifts of CMB last-scattering and matter-radiation equality. This plot indicates that changes to the energy density in the decade of redshift before recombination would generically have a significantly different impact on these two scales.}
    \label{fig:horizons}
\end{figure}

\section{New physics effects on sound horizon and equality scales}
\label{section:newPhysics}

The method we have outlined above can provide constraints on the Hubble constant which derive from the (projected) scale of matter radiation equality, $k_{\rm eq}$, instead of the sound horizon scale $r_s$. Before proceeding to detailed forecasts, we will briefly justify why new physics models that are proposed to resolve the Hubble tension can differently affect equality and sound horizon scales, making our measurement a useful probe of new physics.

First, we note that equality and sound horizon scales are sensitive to changes in the expansion rate at substantially different redshifts. To illustrate this, we show in Fig.~\ref{fig:horizons} a comparison of how the sound horizon and the horizon at matter radiation equality (assuming this coincides with the Jeans scale) vary with a change in the Hubble rate at a certain redshift; in particular, we plot $\frac{d \ln r_s}{d \ln H(z)}$ and $\frac{d \ln d_{\rm eq}}{d \ln H(z)}$, where $d_{\rm eq}$ is the horizon size at MRE. It can be seen that the the sound horizon scale is substantially affected by changes to the expansion rate at redshifts just above $z\approx 1100$, whereas it only has a weak sensitivity to changes at $z>3400$. In contrast, the sensitivity of the equality scale peaks near $z\approx 3400$ and extends to higher redshift.

Second, we note that many of the most commonly considered new physics models to resolve the Hubble tension (ones that are still consistent with high-precision measurements of the CMB power spectrum) involve a change in the expansion rate in the decade of scale factor just before recombination \citep{Knox:2020}. For example, \cite{Smith:2020} invokes a $10\%$ contribution of early dark energy (EDE) density at $z\sim 3500$ that then dilutes away rapidly. (It has been pointed out by some authors that EDE appears to also cause inconsistencies with large scale structure, at least when the neutrino mass is not marginalized over, see \citealt{Hill:2020, Ivanov:2020b, dAmico:2020b} .  \citet{Chudaykin:EDE}, however, find that EDE can improve the goodness of fit to a combination of {\it Planck}, SPTpol, and large scale structure data.) While any details will, of course, be model-dependent, the different redshift sensitivity in Fig.~\ref{fig:horizons} suggests that probes of the expansion rate or horizon size at $z\sim 1100$ and $z \sim 3400$ could generically give results that differ at a significant and detectable level.

A measurement of the Hubble constant derived from the equality scale that differs from measurements derived using $r_s$ would therefore give evidence for the hypothesis that new physics is responsible for tensions in the Hubble constant measurements; by constraining the redshift range over which the expansion rate is modified, such a result would also give further insight into the phenomenology of any new physics.  In contrast, agreement of Hubble measurements derived from both the sound horizon and equality scales at high precision would provide a valuable test of the $\Lambda$CDM framework at high redshifts.

\section{Forecasting Methodology and Datasets}
\label{sec:forecast_setup}

Below, we will forecast constraints on $H_0$ for future experiments, as well as present results from current data.  We now discuss our forecasting methodology and the relevant datasets.  Our forecasts are derived from simulated likelihood analyses in which we generate and then analyze mock CMB lensing power spectrum measurements.  The mock measurements are noiseless so that the likelihood necessarily peaks at the input parameter values; the covariance assumed when analyzing the mock data, however, includes realistic noise levels so that the shapes and sizes of the parameter contours are meaningful.  We use the \texttt{MultiNest} \citep{multinest} algorithm in \texttt{CosmoSIS} \citep{cosmosis} to generate parameter samples, and use the \texttt{getdist} \citep{getdist} package to generate contour plots.  The matter power spectrum is computed using \texttt{CAMB} \citep{CAMB}.  We vary the parameters $h$, $n_s$, $A_s$, $\Omega_m$, $\Omega_b$, and $\Omega_{\nu} h^2$ in our analysis.  

\subsection{CMB lensing}
\label{sec:cmblensing_survey}

For our analysis of current data, we use the CMB lensing measurements from \citet{Plancklensing2018} and the corresponding parameter chains made available on the Planck Legacy Archive.\footnote{\url{https://pla.esac.esa.int/}}

For the forecasts, we assume a Gaussian likelihood and covariance for the observed CMB lensing power spectrum, $\hat{C}^{\kappa\kappa}(L)$. This is a good approximation because CMB lensing probes mainly large, near-linear scales in the matter distribution and the reconstruction noise becomes close to Gaussian when averaged in bandpowers (e.g., \citealt{HuOkamoto}). 

In this Gaussian limit, the covariance is given by
\begin{eqnarray}
\mathrm{cov}(\hat{C}^{\kappa\kappa}(L),\hat{C}^{\kappa\kappa}(L')) = \frac{\delta_{LL'}2\left( C^{\kappa\kappa}(L) + N^{\kappa\kappa}(L) \right)^2}{f_{\rm sky} (2 L+1)},  \nonumber \\
\end{eqnarray}
where $C^{\kappa\kappa}(L)$ is the model CMB lensing power spectrum and $N^{\kappa\kappa}(L)$ is the noise power spectrum.  We will consider the CMB-S4 experiment – a low-noise, ground-based CMB polarization survey which will be deployed in the second half of this decade – as an example for the source of future CMB lensing data in our forecasts. In particular, we will assume the $f_{\rm sky}=0.4$ wide legacy survey, with a lensing reconstruction noise forecast as in \cite{S4DSR}.  We set $L_{\rm min} = 20$,  $L_{\rm max} = 3000$. 

\subsection{CMB-motivated \texorpdfstring{$A_s$}{Lg} constraint}

As discussed in \S\ref{sec:parameter_discussion}, we use a conservative prior on $A_s$ to break the $L_{\rm eq}$-$A_s$ degeneracy.  We endeavor to obtain this constraint on $A_s$ without making assumptions about $r_s$.

The CMB temperature power spectrum constrains $A_s$ to high precision \citep{Planck:parameters}.  The dominant limiting factor in any measurement of $A_s$ from the CMB is a measurement of the optical depth $\tau$, which determines how much the CMB power spectrum amplitude has been reduced by Thomson scattering of CMB photons on their journey from the last scattering surface. Since the value of $\tau$ is determined from features in the very largest scales of the CMB power spectra (in particular the polarized CMB spectrum $EE$), much larger than the sound horizon scale, its inference should be unaffected by changes in $r_s$. Due to this (and since $A_s$ impacts the amplitude of the spectrum, while $r_s$ changes its angular scale) one would not expect a very strong degeneracy between $A_s$ and $r_s$. Indeed, no significant degeneracy between $r_s$ and  $A_s$ is  seen in the {\it Planck} parameter chains \citep{Planck:parameters}. This suggests that we may safely include a CMB-inspired prior on $A_s$ without significant concerns about making any implicit assumptions about $r_s$. 

Motivated by current constraints, for our forecasts we will adopt a conservative 4\% prior on $A_s$: $A_s = (2.19 \pm 0.09)\times 10^{-9}$.  In comparison, \citet{Planck:parameters} constrains $A_s$ to roughly the 1.6\% level.  We will show that tightening the $A_s$ prior below 4\% does not significantly improve the forecast $H_0$ constraints.  For current data, our analysis is even less sensitive to the $A_s$ prior (see discussion in \S\ref{sec:constraint_explanation}); to be maximally conservative, we will therefore use an even looser 8\% prior for the analysis of current data.

\subsection{Supernovae}
\label{sec:supernovae}

As discussed in \S\ref{sec:parameter_discussion}, our constraints on $H_0$ from the CMB lensing power spectrum rely on external measurements of $\Omega_m$ that are independent of $r_s$.  To this end, we consider constraints on $\Omega_m$ from observations of supernovae.  The supernovae measurements alone do not constrain $H_0$ since the absolute luminosity of the supernovae is not known without the supernovae measurements being tied to the CDL.  The shape of the supernovae redshift-apparent brightness relation, however, constrains $\Omega_m$ without any reliance on $r_s$.  

For the analysis of current data, we consider supernova constraints from the Pantheon sample \citep{Scolnic:2018}.  This analysis found $\Omega_m = 0.298 \pm 0.022$ in a flat $\Lambda$CDM fit to  supernovae data alone (including systematic errors).  We will take the approach of simply using the marginalized posterior from \citet{Scolnic:2018} on $\Omega_m$ as a prior in our analysis of current CMB lensing measurements (see \S\ref{sec:current_data}). 

For our forecast constraints, we adopt $\sigma({\Omega_m}) = 0.012$, where we use the notation $\sigma(x)$ to denote the $1\sigma$ uncertainty on $x$.  Very roughly, this level of uncertainty could be obtained by doubling the number of supernova in the Pantheon sample, combined with a factor of two reduction in systematic uncertainty.  Put another way, $\sigma(\Omega_m) = 0.012$ is roughly consistent with the forecast for the analysis of 10,000 photometric supernova from  the Vera Rubin Observatory Legacy Survey of Space and Time (LSST, \citealt{LSSTsciencebook}), assuming that there is no systematic floor in the supernovae distance determinations.  Since LSST is expected to obtain roughly 50,000 supernovae per year, depending on the systematic floor, our forecast may therefore be viewed as conservative.  We emphasize that our intent here is to illustrate our ideas and obtain a rough estimate of the $H_0$ constraints that can be achieved with our method, rather than to perform the most realistic future forecast possible.

\begin{table}
    \centering
    \begin{tabular}{|c|c|c|}
    \hline
        Parameter &  Fiducial value & Prior  \\
    \hline
        $h$ & $0.69$ & $[0.4, 0.9]$ \\
        $n_s$ & $0.97$ & $[0.87, 1.07]$ \\
        $A_s$ & $2.19\times 10^{-9}$ & $[0.5, 5.0]\times 10^{-9}$ \\
        $\Omega_m$ & $0.3$ & $[0.1, 0.9]$ \\
        $\Omega_b$ & $0.048$ & $[0.03, 0.077]$ \\
        $\Omega_{\nu}h^2$ & $0.00083$ & $[0.0006, 0.01]$ \\  \hline
        $\Omega_b h^2$ &  0.02285 & $[0.014, 0.032]$ \\ \hline
    \end{tabular}
    \caption{Fiducial parameter values and priors used in the forecasts.   Although we do not vary $\Omega_b h^2$ independently, we impose a conservative prior on this parameter combination so as to not explore highly unphysical regions of parameter space. }
    \label{tab:priors}
\end{table}

\begin{figure*}
    \centering
    \includegraphics[scale=0.5]{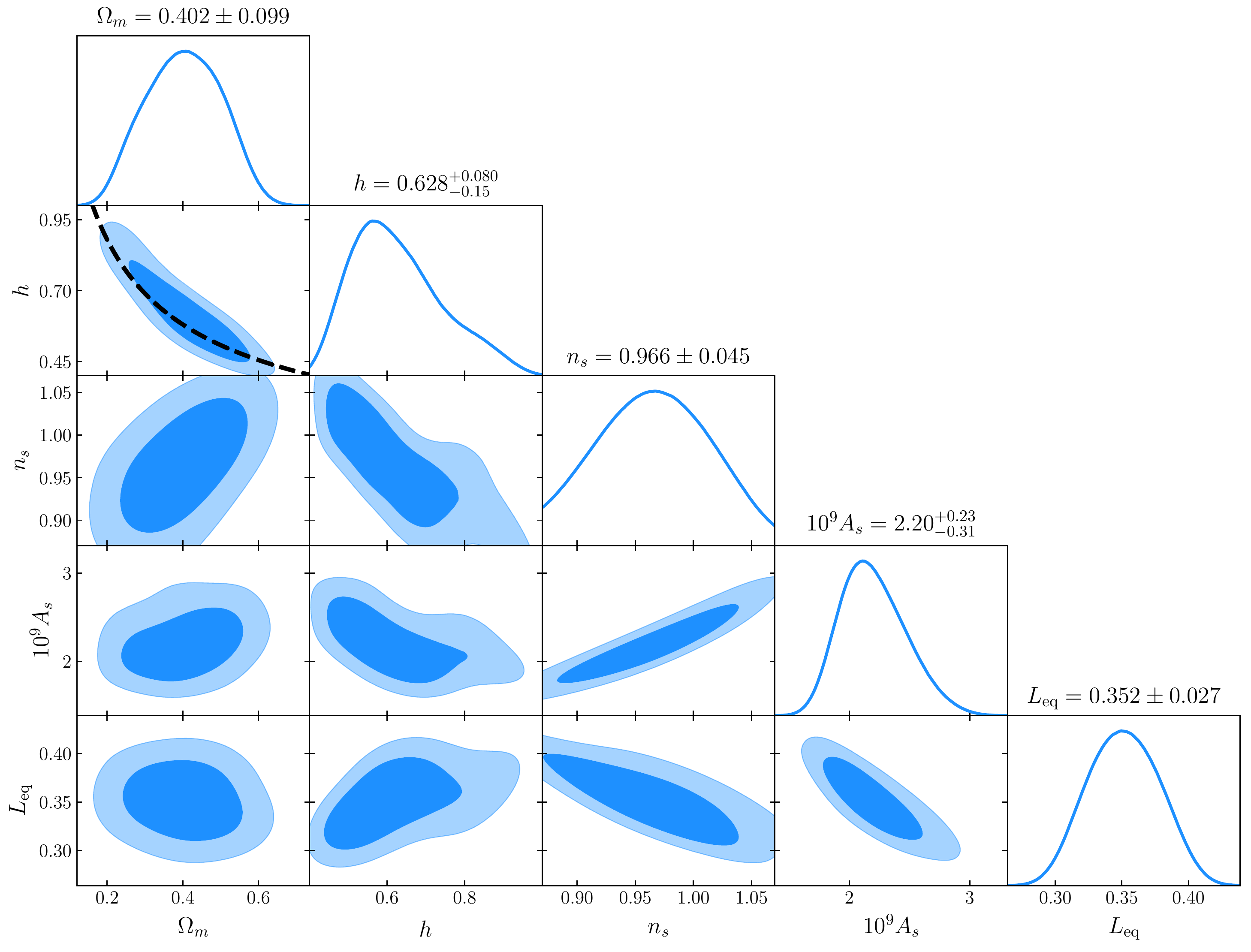}
    \caption{Forecast parameter constraints from CMB-S4-like measurements of the CMB lensing power spectrum.  Information about $H_0$ enters via $L_{\rm eq} \equiv k_{\rm eq}\chi_* \propto  \Omega_m^{0.6} h$; the black dashed curve shows the expected $h \propto \Omega_m^{-0.6}$ degeneracy direction.  Degeneracy between $L_{\rm eq}$ and $A_s$ limits the ability of the CMB lensing power spectrum to constrain $L_{\rm eq}$, and thus $H_0$.  We have marginalized over $\Omega_{\nu}h^2$ with the priors from Table~\ref{tab:priors} when generating this figure; both parameters are prior dominated. }
    \label{fig:no_priors}
\end{figure*}

\subsection{Galaxy lensing}

Galaxy lensing provides an alternative to supernovae for obtaining a constraint on $\Omega_m$.  In cases where we make use of galaxy lensing, our forecast is designed to roughly correspond to LSST \citep{LSSTsciencebook}.  We adopt $f_{\rm sky} = 0.5$, and source number densities of  $2\,{\rm arcmin}^{-2}$ in four redshift bins between $z = 0$ and $z = 1.2$.  We assume shape noise with $\sigma_{\epsilon} = 0.3$.  We use \texttt{CosmoSIS} \citep{cosmosis} to generate generate the simulated data and covariance matrix, ignoring the contributions of three-point and higher-order correlators to the covariance matrix.  For current data, these higher order contributions generally make a negligible contribution to the weak lensing covariance matrix.  For future datasets such as LSST, however, this is not the case.  Our forecast constraints on $\Omega_m$ from galaxy lensing may therefore be viewed as optimistic.

\subsection{Priors}
\label{sec:priors}

Our choice of priors for the forecasts is summarized in Table~\ref{tab:priors}.  This choice roughly follows that of the weak lensing analysis presented in \citet{Troxel:2018}.  For $h$, $n_s$, $A_s$ and $\Omega_m$, our priors are not informative.  For $\Omega_b$ and $\Omega_{\nu} h^2$, the priors are weakly informative.  Our prior on $\Omega_b$ is more than an order of magnitude looser than the constraints from \citet{Planck:parameters}.  Following \citet{DESy1}, we use a prior on the density parameter of neutrinos of $\Omega_{\nu} h^2 \in [0.0006, 0.01]$.  This roughly corresponds to a prior on the sum of the neutrino masses of $\sum m_{\nu} \in [0.06,1.0]  \,{\rm eV}$, where the lower limit is approximately the minimum allowed by neutrino oscillation experiments \citep{Fogli:2008}.  The upper limit of this prior is comparable to recent limits from the KATRIN experiment \citep{KATRIN}, and is significantly above the level of current constraints from a combination of CMB and BAO data \citep{Planck:parameters}.  We additionally impose a very conservative (roughly 40\%) flat prior on the parameter combination $\Omega_b h^2$.  This prior ensures that we do not explore highly unphysical regions of parameter space for current data (where constraints are quite weak), as discussed in more detail below. However, in general, this prior has a small impact on our forecast results.

\section{Forecast Constraints on \texorpdfstring{$H_0$}{Lg} without the Sound Horizon Scale}
\label{sec:results}

We now forecast constraints on the Hubble constant that can be obtained independently of the sound horizon information, using a combination of CMB lensing measurements and other data.  We begin by considering forecasts for CMB-S4.  We present results from current data in \S\ref{sec:current_data}.

Fig.~\ref{fig:no_priors} shows the $\Lambda$CDM parameter constraints for CMB lensing measurements of a CMB-S4-like survey.    The constraint obtained on $H_0$ in this case is weak because of the large degeneracies with other parameters.  The constraints on $L_{\rm eq}$ (from which the $H_0$ constraint derives) are degraded by degeneracies with $A_s$ and $n_s$.  This is not surprising given Eq.~\ref{eq:cmb_limber_tfapprox} and Eq.~\ref{eq:cmb_limber_ns1}.   Second, the uncertainty on $\Omega_m$ is large, so it would not be possible to constrain $H_0$ very tightly even if $L_{\rm eq}$ were measured to high precision.  Roughly, since $L_{\rm eq} \propto h \Omega_m^{0.6}$, we need to know $\Omega_m$ to roughly $8\%$ precision to have any hope of constraining $H_0$ to the 5\% level with a  measurement of $L_{\rm eq}$.

\begin{figure*}
    \centering
    \includegraphics[scale=0.5]{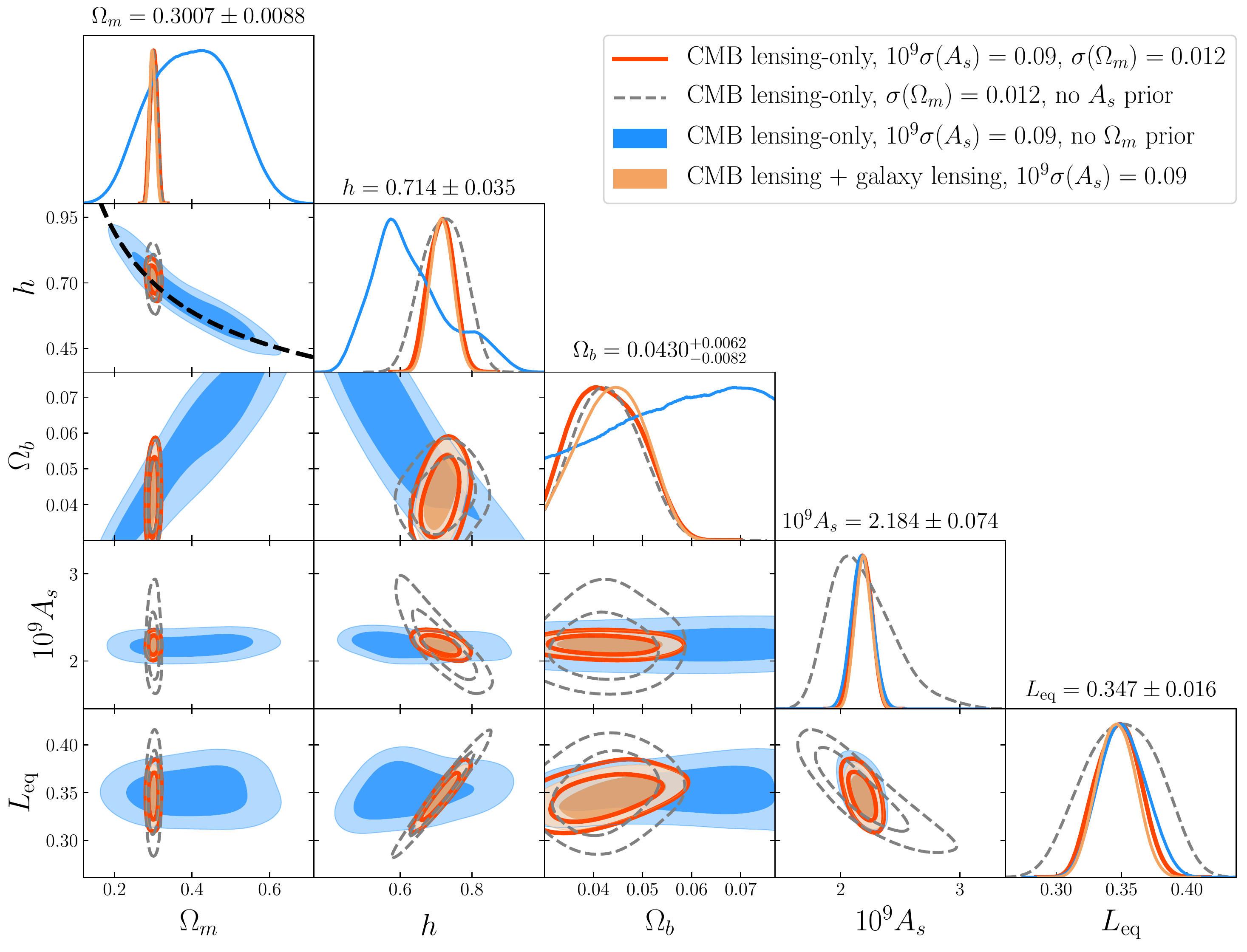}
    \caption{Forecast constraints for a CMB-S4 measurement of the CMB lensing power spectrum upon including external $A_s$ and $\Omega_m$ information.   By imposing a conservative, CMB-motivated (but $r_s$ independent) prior on $A_s$ we break the $A_s$-$L_{\rm eq}$ degeneracy, leading to a constraint on $L_{\rm eq} \equiv k_{\rm eq} \chi_{*}$ (blue curves; compare to Fig.~\ref{fig:no_priors}).  Since $L_{\rm eq} \propto \Omega_m^{0.6} h$, the resultant constraints on $\Omega_m$ and $H_0$ are degenerate (the black dashed curve illustrates the expected degeneracy).  By including an additional $\Omega_m$ constraint from supernovae we break the $\Omega_m$-$H_0$ degeneracy, and obtain a constraint on $H_0$ (red curve, with marginalized parameter uncertainties reported along the diagonal).  As an alternative to a supernova constraint on $\Omega_m$, we also consider adding information from an LSST-like measurement of cosmic shear (orange curve); in combination with CMB lensing, galaxy lensing also provides $\Omega_m$ information which serves to break the $H_0$-$\Omega_m$ degeneracy.  Because of large-scale information in the CMB lensing power spectrum, our constraints are not very sensitive to the $A_s$ prior.  If this prior is completely removed, we still obtain a constraint on $H_0$ (grey dashed curve).
    }
    \label{fig:As_prior}
\end{figure*}

\subsection{CMB lensing with \texorpdfstring{$A_s$}{Lg} prior and supernova \texorpdfstring{$\Omega_m$}{Lg} constraint}
\label{sec:main_results}

With a conservative external $A_s$ prior, the $L_{\rm eq}$--$A_s$ degeneracy can be broken, leading to a tight constraint on $L_{\rm eq}$.  Since $L_{\rm eq}$ depends only on $\Omega_m$ and $H_0$, given an additional $\Omega_m$ constraint, $H_0$ can then be determined. As described in \S\ref{sec:forecast_setup}, our conservative $A_s$ prior ($10^{9}A_s = 2.19 \pm 0.09$) is motivated by current constraints from the primary CMB, while our $\Omega_m$ constraint ($\Omega_m = 0.3 \pm 0.012$) is forecast for future supernovae measurements.  Both the $A_s$ and $\Omega_m$ constraints are independent of assumptions about $r_s$.  

In Fig.~\ref{fig:As_prior} we show (blue curves) the results of imposing the CMB-motivated $A_s$ prior on the CMB lensing constraints from  Fig.~\ref{fig:no_priors}.  Imposing this prior results in tighter constraints on $L_{\rm eq}$ than in Fig.~\ref{fig:no_priors}, as expected.  The results of additionally imposing the supernova prior on $\Omega_m$ are shown with the red curves in Fig.~\ref{fig:As_prior}.  We find that an uncertainty on $H_0$ of 3.5 km/s/Mpc is achieved.  We will refer to this as our pessimistic forecast, since it assumes a conservative $A_s$ prior and a modestly improved $\Omega_m$ from future surveys.

We note that the constraints on $H_0$ are not very sensitive to the $A_s$ prior.  As can be seen in Fig.~\ref{fig:no_priors}, even in the absence of any $A_s$ prior, we obtain some constraint on $L_{\rm eq}$.  This is likely because low-$L$ information in the lensing power spectrum breaks the degeneracy between $A_s$ and $L_{\rm eq}$.  In the absence of the $A_s$ prior, the combination of CMB lensing and the $\Omega_m$ constraint yields an $H_0$ uncertainty of roughly 6 km/s/Mpc (grey dashed curves in Fig.~\ref{fig:As_prior}). 

It is also interesting to consider how the constraints on $H_0$ could improve with more optimistic constraints on $A_s$, $\Omega_m$, and $\Omega_b$.  For CMB-S4 noise levels, tightening the prior on $A_s$ alone by a factor of three yields an $H_0$ uncertainty of $3.1$~km/s/Mpc, only a modest improvement relative to the $3.5$~km/s/Mpc uncertainty obtained with the pessimistic forecast. Tightening the $\Omega_m$ prior by a factor of three to $\sigma({\Omega_m}) = 0.004$ would improve the $H_0$ constraint only marginally to $\sigma({H_0}) = 3.3$~km/s/Mpc.  With combined priors of  $10^9 \sigma({A_s}) = 0.03$ and $\sigma({\Omega_m}) = 0.004$, we find $\sigma({H_0}) = 2.9$~km/s/Mpc.  We refer to this forecast as our optimistic forecast. 

Our analysis also imposes conservative priors on $\Omega_b$ and the sum of the neutrino masses.  Tightening the fiducial 40\% prior on $\Omega_b$ to the 10\% level  would slightly improve the $H_0$ constraint to $\sigma({H_0}) = 3.3$~km/s/Mpc, as there is a weak degeneracy between $\Omega_b$ and $H_0$ (see Fig.~\ref{fig:As_prior}).  This degeneracy presumably results from the baryonic suppression of the matter power spectrum at small scales, which is somewhat degenerate with the impact of changing $L_{\rm eq}$. Tightening the prior on neutrino mass has little impact on the $H_0$ constraint. 

It is also interesting to consider how the $H_0$ constraints would improve with even more futuristic CMB data.  We find that reducing $N_L^{\kappa\kappa}$ significantly below the CMB-S4 level does not result in dramatic improvements to the constraints on $H_0$ because of cosmic variance from the already signal-dominated large scale CMB lensing power spectrum, and because of the of the interplay between various parameter degeneracies.  For a zero-noise CMB lensing measurement (with $L_{\rm min} = 20$ and $L_{\rm max} = 3000$), $\sigma({\Omega_m}) = 0.006$ and $10^9 \sigma({A_s}) = 0.03$, we find $\sigma({H_0}) = 2.2$~km/s/Mpc. 

\subsection{Alternative dataset and prior choices}

We now consider two alternative combinations of datasets and priors that might provide similar constraints on the Hubble constant.

First, instead of relying on supernovae to constrain $\Omega_m$, one can use measurements of the galaxy lensing power spectrum.   The combination of galaxy lensing measurements with CMB lensing measurements breaks the  $\sigma_8$-$\Omega_m$ degeneracy, leading to a tight constraint on $\Omega_m$.  This constraint on $\Omega_m$ then breaks the $H_0$-$\Omega_m$ degeneracy.  This possibility is shown as the orange curve in Fig.~\ref{fig:As_prior}; we find a Hubble constraint of $\sigma(H_0) = 3.1$~km/s/Mpc.  Note that, for simplicity, we assume no covariance between the galaxy and CMB lensing measurements, as would be appropriate for non-overlapping surveys.  Even in the case of overlapping surveys, though, the covariance between the galaxy and CMB lensing measurements is reduced by the different redshift sensitivities of the two observables.  

Second, rather than imposing an $A_s$ prior to break the $L_{\rm eq}$-$A_s$ degeneracy, an alternate possibility is to impose an $n_s$ prior.  As seen in Fig.~\ref{fig:no_priors}, $n_s$, $A_s$ and $L_{\rm eq}$ are all degenerate, which is not surprising from Eq.~\ref{eq:cmb_limber_tfapprox}.  By imposing a prior on $n_s$, we can break this degeneracy and obtain tight constraints on $L_{\rm eq}$ which can again be used to constrain $H_0$ in combination with an $\Omega_m$ prior. 

As with $A_s$, the  CMB temperature power spectrum constrains $n_s$.  However, unlike  $A_s$, the $n_s$ constraint from the CMB is degenerate with $r_s$.  An $r_s$-independent constraint on $n_s$ could instead perhaps be obtained from Lyman-$\alpha$ forest constraints on the matter power spectrum.  Although this possibility certainly requires further exploration, our general motivation is that the amplitude of broadband structure on small scales (and hence $n_s$) could potentially be determined in a way that is not degenerate with the BAO features and hence $r_s$.  To test the effectiveness of this approach, we adopt a prior of $n_s = 0.96 \pm 0.015$.  This prior is conservative compared to current constraints from \citet{Planck:parameters}.  With this $n_s$ prior and the $\sigma(\Omega_m) = 0.012$ constraint, the resultant $H_0$ constraint is $68.84^{+3.4}_{-4.1}$.   Since this constraint is weaker than our pessimistic forecast using an $A_s$ prior, we do not pursue this possibility further.

\subsection{Verifying independence of the \texorpdfstring{$H_0$}{Lg} constraints from sound horizon physics}
\label{sec:sound_horizon_info}

Our goal in this analysis is to constrain $H_0$ without relying on assumptions about the sound horizon scale.  As noted previously, the sound horizon scale is imprinted on the matter power spectrum in two ways.  First, the sound horizon enters via the baryon acoustic oscillations.  These, however, are washed out by the line-of-sight integration in Eq.~\ref{eq:limber}.  Secondly, power below $r_s$ is suppressed because the baryons cannot cluster below the sound horizon scale prior to recombination.  While the baryon suppression feature is also washed out by the line-of-sight integration, it is still visible to some extent in the CMB lensing power spectrum.   It is conceivable, then, that the CMB lensing measurements of this suppression could be used to determine the angular scale of $r_s$.   Since the amplitude of the baryonic suppression depends on $\Omega_b h^2$, measurement of this suppression could also calibrate the $r_s$ ruler, leading to an $H_0$ constraint. Indeed, this mechanism may make a significant contribution to the forecast constraints on the broadband power spectrum from Euclid presented by \cite{Chudaykin:2019}.  This possibility is somewhat worrying for our analysis since it could mean that our Hubble constraints are actually not independent of $r_s$.

The fact that our arguments about breaking parameter degeneracies (which ignored the impact of the sound horizon scale) work as expected suggests that information about $H_0$ is indeed entering via $L_{\rm eq}$ rather than $r_s$.  As shown by the black dashed curve in Fig.~\ref{fig:As_prior}, the $H_0$ information closely follows the expected $L_{\rm eq}$ degeneracy direction. 

Furthermore, if the information on $H_0$ were entering via the baryon suppression feature, then we would expect the constraints on $H_0$ to get weaker with reduced $\Omega_b$ since smaller $\Omega_b$ means less baryon suppression.  When we repeat our analysis using a much lower value of $\Omega_b = 0.004$ (setting $\Omega_b$ much lower than this causes \texttt{CAMB} to crash) we find that the constraints on $H_0$ are actually {\it improved} by roughly 35\% relative to the fiducial analysis with $\Omega_b = 0.048$.  This implies that the baryon suppression is not contributing to our $H_0$ constraints, and suggests that the main impact of the baryon suppression feature is actually to {\it degrade} our $H_0$ constraints because of weak degeneracies with other parameters.

To further increase our confidence that the sound horizon scale is not influencing our constraints, we explore how the CMB lensing power spectrum constrains $H_0$ using a Fisher matrix analysis.  To this end, we generate a mock power spectrum using a combination of the dark matter only transfer function from \citet{BBKS} and a toy model for the impact of baryon suppression.  We refer to the wavenumber corresponding to the sound horizon scale as $k_s$, i.e. $k_s \sim 1/r_s$.  At small scales, $k \gg k_s$, the baryons are pressure supported and unable to cluster.  Consequently, the fraction of matter that clusters is reduced by a factor $1- \Omega_b/\Omega_m$, and the transfer function is similarly reduced.  At large scales, $k \ll k_s$, the baryons act like cold dark matter and the transfer function is unaffected by the baryons.  In our toy model for the impact of baryons, we assume that the transition between these two regimes is a step function.  In reality, of course, the transition is not infinitely sharp and is accompanied by additional features due to acoustic oscillations \citep{EH:1998}.  By exaggerating the sharpness of the baryon suppression, we are enhancing the information in this feature, and our analysis is therefore conservative with respect to the information about $H_0$ that this feature contributes.

\begin{figure}
    \centering
    \includegraphics[scale=0.58]{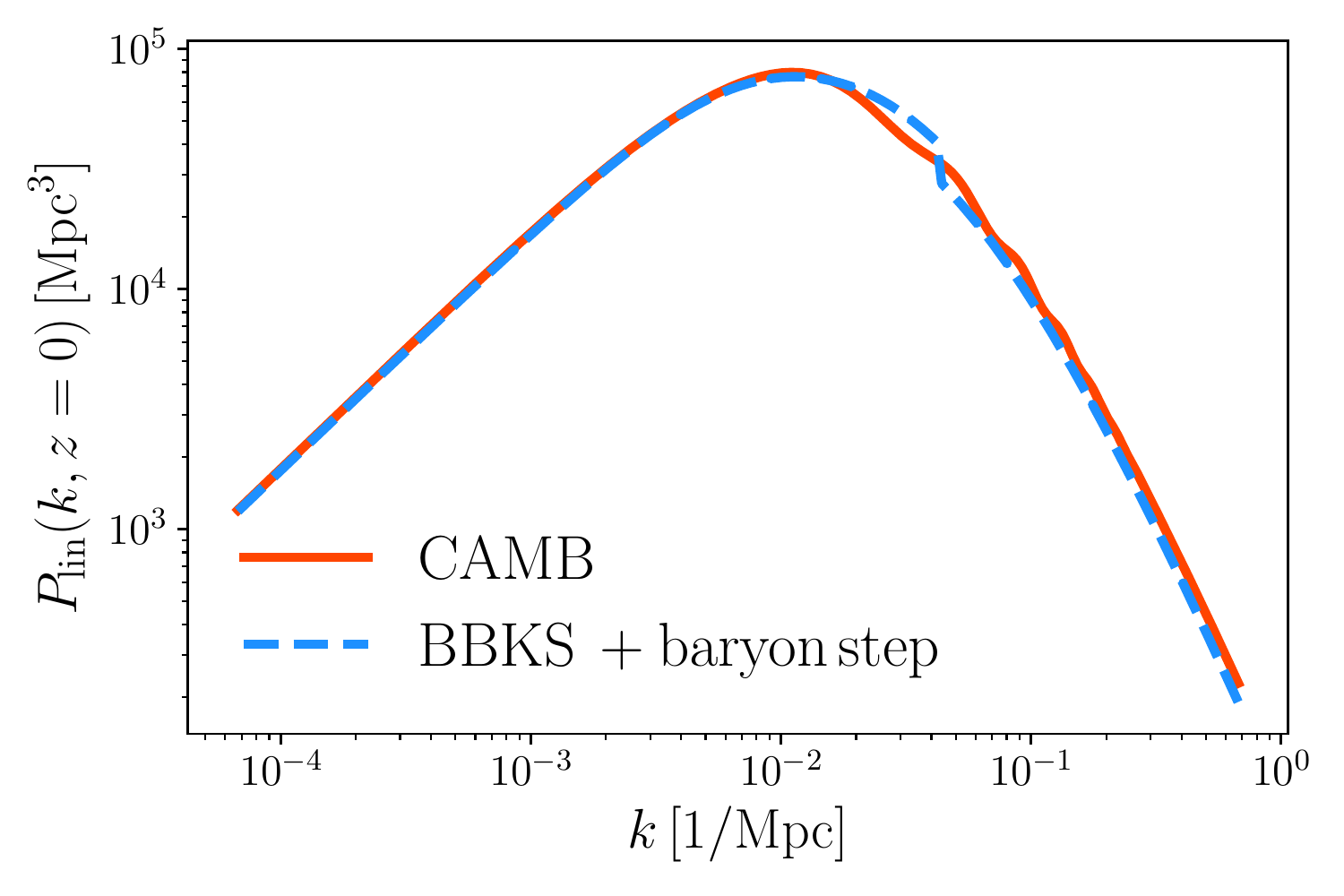}
    \caption{The linear matter power spectrum computed with \texttt{CAMB}  (red solid) and with the toy model of Eq.~\ref{eq:toy_pk} (blue dashed).  In this toy model, we adopt the BBKS transfer function \citep{BBKS}, and model the impact of baryons on the matter power with a step-like suppression on scales smaller than $r_s$.  We use this toy model to explore how the CMB lensing power spectrum constrains $H_0$, and to confirm that our constraints are not informed by information about the sound horizon scale, $r_s$.}
    \label{fig:toy_model}
\end{figure}

Explicitly, our toy model for the power spectrum is:
\begin{multline}
\label{eq:toy_pk}
    P^{\rm toy}(k,z) \propto \frac{1}{(\Omega_m H_0^2)^2} A_s k^{n_s} a^2(z) G^2(z) \\
    T_{\rm BBKS}^2(k;k_{\rm eq}) \left[ 1 - (\Omega_b/\Omega_m)\Theta(k - k_s) \right]^2,
\end{multline}
where $T_{\rm BBKS}(k;k_{\rm eq})$ is the transfer function from \citet{BBKS} which is only a function of $k_{\rm eq}$, and $\Theta(k)$ is a step function.  We show a comparison of our mock power spectrum model to that computed from \texttt{CAMB} \citep{CAMB} in Fig.~\ref{fig:toy_model}.  We calculate the corresponding CMB lensing power spectrum using the Limber approximation from Eq.~\ref{eq:limber}, substituting our toy model for the matter power spectrum.

To explore how information about $r_s$ enters into the $H_0$ constraints, we introduce a new parameter $\alpha_{s}$ which takes $k_s \rightarrow \alpha_s k_s$.  In effect, this parameter allows the sound horizon scale to depart from its standard value.  We then compute the Fisher information matrix, $F_{ij}$, where $i,j  \in \{ \Omega_m, A_s, H_0, \Omega_b, \alpha_s \}$.  The forecast errors on $H_0$ are given by $\sigma({H_0}) = [\mathbf{F}^{-1}]_{H_0 H_0}^{1/2}$. We impose priors of $\sigma(\Omega_m) = 0.012$, $\sigma(A_s) = 0.04 A_s$, and $\sigma(\Omega_b) = 0.2$, matching our prior choice for the likelihood analysis (except for $\Omega_b$, for which we impose a top-hat prior in the likelihood analysis).

When we fix $\alpha_s = 1$, we find that the constraint on $H_0$ projected for CMB-S4 noise levels is $\sigma({H_0}) = 2.18$~km/s/Mpc, which agrees fairly well with the constraints obtained from the full likelihood analysis described in \S\ref{sec:main_results}.  Note that we do not expect this constraint to agree perfectly with the likelihood analysis because our Fisher analysis uses a toy power spectrum model, because we do not vary $n_s$ or $\Omega_{\nu} h^2$, and because we impose a Gaussian (rather than top-hat) prior on $\Omega_b$.  

\begin{figure*}
    \centering
    \includegraphics[scale=0.45]{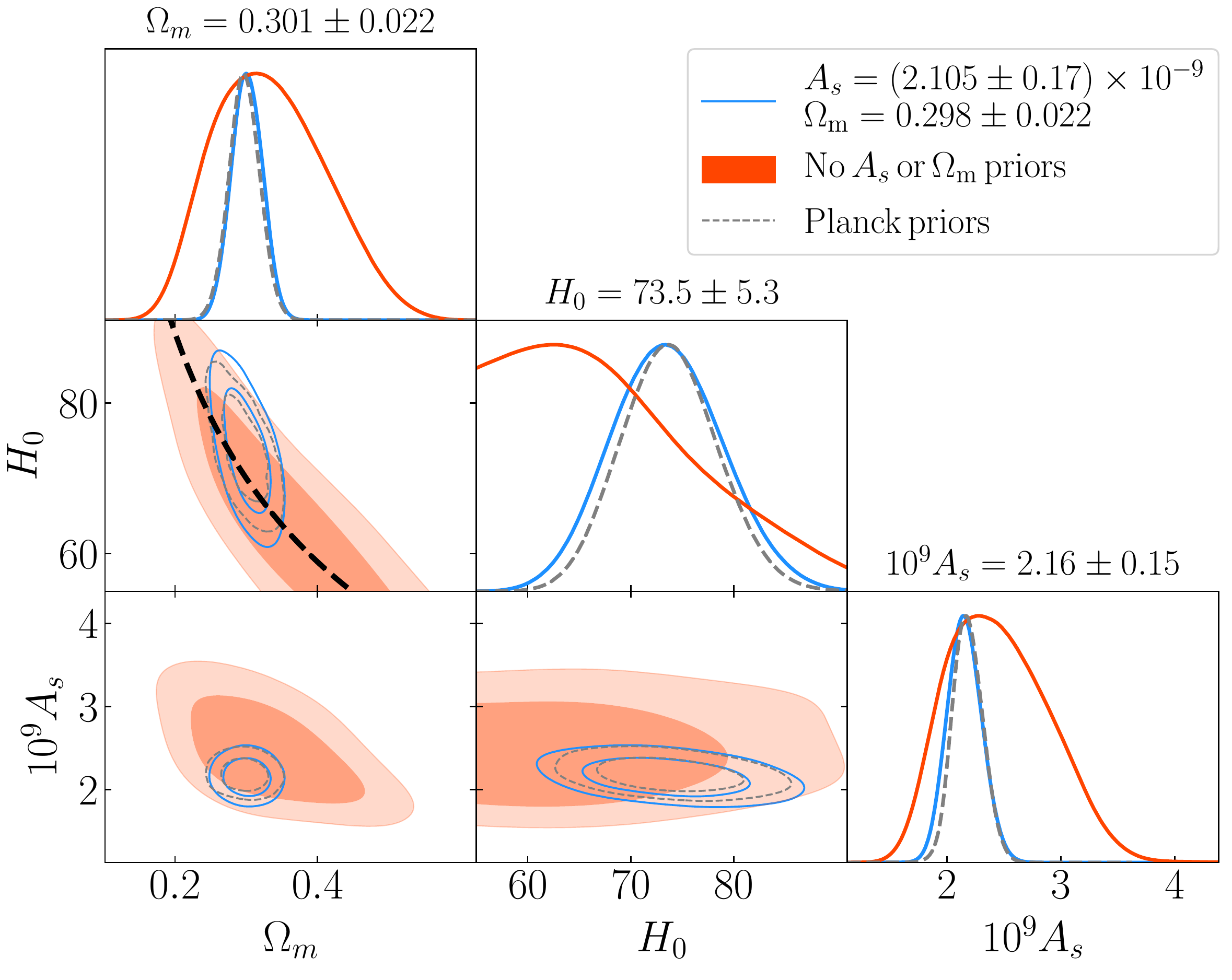}
    \caption{Constraints on $H_0$ from current {\it Planck} CMB lensing measurements \citep{Plancklensing2015} before (orange) and after (blue) imposing the supernova $\Omega_m$ constraint \citep{Scolnic:2018} and the conservative, CMB-motivated (but $r_s$ independent) $A_s$ prior. The resultant constraint of $H_0 = 73.5\pm5.3$~km/s/Mpc is independent of the sound horizon scale, $r_s$.  The black dashed curve illustrates the expected degeneracy between $h$ and $\Omega_m$ when information about $H_0$ enters via $L_{\rm eq} \equiv k_{\rm eq} \chi_*$.  The grey dashed curves indicate the result obtained with an alternate choice of priors, see discussion in text. }
    \label{fig:current_data}
\end{figure*}

We next allow $\alpha_s$ to vary, and marginalize over this quantity after imposing a broad Gaussian prior with $\sigma({\alpha_s}) = 0.25$ to avoid considering models that are drastically inconsistent with current cosmological observations.  Allowing $\alpha_s$ to vary effectively erases information about the absolute calibration of $r_s$.  In this case, we find that the error on $H_0$ becomes $\sigma({H_0}) = 2.34$~km/s/Mpc, i.e. an increase in the uncertainty of only 7\%.  The small change in the uncertainty on $H_0$ suggests that $r_s$ has essentially no impact on our $H_0$ constraints.  Note that some degradation of the $H_0$ constraint is expected even if no information about $r_s$ is used to constrain $H_0$, since the impact of changing $r_s$ will be somewhat degenerate with the impact of changing $k_{\rm eq}$.  

We can also use the Fisher analysis to illustrate how $k_{\rm eq}$ informs the $H_0$ constraint.  Similar to $\alpha_s$, we introduce a new parameter, $\alpha_{\rm eq}$, which scales $k_{\rm eq}$ in the Fisher calculation.  We find that allowing $\alpha_{\rm eq}$ to vary degrades the $H_0$ constraint to $\sigma({H_0}) = 12.6$~km/s/Mpc, i.e. a factor of 5.8 increase in the uncertainty. This large increase in the uncertainty confirms that nearly all the constraining power on $H_0$ is arising from the (projected) equality scale.

\section{Constraints from current data}
\label{sec:current_data}

\subsection{An \texorpdfstring{$r_s$}{Lg}-independent constraint on \texorpdfstring{$H_0$}{Lg} from CMB lensing and external data}

We now consider how current measurements of the CMB lensing power spectrum from \citet{Plancklensing2018} constrain $H_0$ in combination with external priors. We emphasize that these constraints are obtained without relying on the sound horizon scale $r_s$.  In the interest of being maximally conservative, for current data we degrade the $A_s$ prior to the 8\% level, rather than the 4\% level assumed in the forecasts.

The red curves in Fig.~\ref{fig:current_data} show the constraints from \citet{Plancklensing2018} obtained using the priors in Table~\ref{tab:priors}.  The blue curve results from additionally imposing an $\Omega_m = 0.298 \pm 0.022$ constraint from current supernovae measurements \citep{Scolnic:2018}, and an $A_s = (2.105 \pm 0.17)\times 10^{-9}$ prior (which, as discussed previously, is motivated by CMB observations, but is very conservative and is, in any case, expected to be independent of $r_s$ information).  The  derived constraint on $H_0$ is $H_0 = 73.5 \pm 5.3\,{\rm km}/{\rm s}/{\rm Mpc}$. 

Relative to the forecasts presented previously, the larger uncertainties with current data mean that volume effects in the marginalized posterior are potentially more important, and we expect greater sensitivity to our priors.  Indeed, without the $\Omega_b h^2$ prior, the data explores a region of parameter space at very high, unphysical $\Omega_b h^2$.  Imposing the conservative $\Omega_b h^2$ prior in Table~\ref{tab:priors} prevents this from occurring.  Our results are not very sensitive to this prior: tightening it from roughly 40\% to 5\% changes the recovered value of $H_0$ from $73.5  \pm  5.3$ to $73.6 \pm  4.7$~km/s/Mpc. As an additional test of stability to prior choices, we also consider an alternate prior choice from the analysis of \citet{Plancklensing2018}.  The main difference between the \citet{Plancklensing2018} priors and our fiducial prior choice is that the former additionally imposes constraints of $n_s = 0.96 \pm 0.02$ and $\Omega_b h^2 = 0.0222 \pm 0.0005$.  Since imposing the $A_s$ prior removes most of the degeneracy between $n_s$ and $H_0$, we do not expect the $n_s$ prior to significantly impact the results.  The $\Omega_b h^2$ prior is derived from the BBN-based analysis of D/H measurements in quasar absorption-line systems.  Motivated in this way, this prior is also independent of $r_s$ physics.  The $H_0$ constraints resulting from this alternative prior choice are shown as the grey dashed curves in Fig.~\ref{fig:current_data}:  we find $H_0 = 73.7 \pm 4.5$~km/s/Mpc, which is in good agreement with the constraint obtained using our fiducial choice of priors.  These results suggest that our $H_0$ constraint is not very sensitive to the choice of priors.

As noted previously, our $H_0$ constraints are not very sensitive to the $A_s$ prior because low-$L$ scales in the CMB lensing power spectrum provide shape information that serves to partially break the degeneracy between $A_s$ and $L_{\rm eq}$.  Since a large fraction of the signal-to-noise of the \citet{Plancklensing2018} lensing power spectrum measurements comes from low-$L$, the $H_0$ constraints obtained in Fig.~\ref{fig:current_data} are even less sensitive to the $A_s$ prior than the forecasts in \S\ref{sec:results}.  Halving the width of the $A_s$ prior (i.e. taking it from 8\% to 4\%) has a small effect, changing the $H_0$ constraint to $73.8 \pm 5.1$~km/s/Mpc.  If the $A_s$ prior is made significantly wider, however, we find that for the fiducial prior choice, the posterior becomes sufficiently unconstrained that prior volume effects appear to significantly impact the $H_0$ constraint; we therefore refrain reporting an $H_0$ constraint in this case.  On the other hand, with the {\it Planck} priors (i.e. the tight $\Omega_b h^2$ prior), these volume effects are minimized, and we obtain a constraint on $H_0$ even without the $A_s$ prior: $H_0 = 72.1 \pm 5.0$~km/s/Mpc (compared to $H_0 = 73.7 \pm 4.5$~km/s/Mpc with the $A_s$ prior).

\subsection{Validating independence of \texorpdfstring{$H_0$}{Lg} constraints from \texorpdfstring{$r_s$}{Lg}}

Although constraints from current data are drawn from large scales in the lensing power spectrum and hence should be even less sensitive to $r_s$ than our forecast constraints, we may again test for their independence from sound horizon physics. Repeating the Fisher matrix calculation described in \S\ref{sec:sound_horizon_info} using the {\it Planck} lensing noise spectrum, we find again that $r_s$ contributes minimal information about $H_0$.  Allowing $\alpha_s$ to vary and marginalizing over this parameter changes the uncertainty on $H_0$ by only $0.2\%$.  On the other hand, allowing $\alpha_{\rm eq}$ to vary in the Fisher matrix calculation increases the $H_0$ uncertainty by a factor of five.  In other words, as intended, the current constraints derive only to a negligible extent from $r_s$, and are instead mainly drawn from $k_{\rm eq}$ information. Fig.~\ref{fig:current_data} also shows that the $H_0$-$\Omega_m$ degeneracy direction is consistent with that predicted assuming information about $H_0$ enters via $L_{\rm eq}$ (the black dashed curve).  The supernova $\Omega_m$ prior breaks this degeneracy.

In the case where we impose alternate priors to test stability of our analysis to prior choices (i.e. the grey dashed curve in Fig.~\ref{fig:current_data}), one might worry that the tight $\Omega_b h^2$ prior could calibrate $r_s$, and thereby influence our $H_0$ constraints.  To make sure that this is not the case, we perform an additional test using the Fisher matrix approach where we {\it fix} the physical size of $r_s$. This is maximally conservative: we are assuming that the $r_s$ ruler is calibrated exactly, and that the lensing measurement can then use the apparent size of this ruler to calibrate $H_0$.  Repeating the Fisher matrix calculation, we find that the $H_0$ constraints actually get weaker by roughly 3\%.  This small change confirms that $r_s$ information does not contribute to our constraints.
 
\section{Prospects for further improvement with other probes}
\label{sec:otherProbes}

In \S\ref{sec:main_results} we saw that it is difficult to use the CMB lensing power spectrum measurement of $L_{\rm eq}$ to obtain a constraint on $H_0$ better than $2$~km/s/Mpc, owing to a combination of cosmic variance and parameter degeneracies.  At large scales, where the $A_s$-$L_{\rm eq}$ degeneracy is broken, cosmic variance of the CMB lensing power spectrum computed from a 2D observable limits our ability to constrain $H_0$.  At small scales, cosmic variance is no longer important, but in this regime degeneracies between $A_s$ and $L_{\rm eq}$, as well as the impact of baryons and neutrinos, degrade the $H_0$ constraints.

An alternative to using CMB lensing to constrain $k_{\rm eq}$ is to instead use measurements of the power spectrum from a spectroscopic galaxy survey to constrain $k_{\rm eq}$. One would expect that, at least with future surveys, since the accessible number of modes in a 3D survey can be significantly larger (on all scales) these measurements could potentially provide the highest signal-to-noise constraints on the Hubble constant without the sound horizon. Indeed, even when BAO information is excluded, forecasts indicate that upcoming galaxy surveys can provide powerful probes of $H_0$ \citep{Chudaykin:2019}.

The challenge with such measurements, however, will be to ensure that the constraints on $H_0$ obtained from fitting the galaxy power spectrum are actually free of $r_s$ information. Both the BAO oscillations and the broadband baryonic suppression features in the matter power spectrum are sensitive to $r_s$, and (unlike in the case of the CMB lensing power spectrum) both may be easily detected in future measurements of the 3D galaxy power spectrum.  Given the parameter sensitivity of the full shape of the matter power spectrum, the $r_s$ ruler can be calibrated to some extent, leading to $H_0$ constraints from the measurements of either the BAO scale or the baryon suppression scale. Indeed, recent measurements of $H_0$ from the full shape of the matter power spectrum measured with BOSS \citep{Philcox:2020,dAmico:2020a, Ivanov:2020a} obtain much of their information about $H_0$ through the BAO scale, and they may also have some degree of sensitivity to the baryon suppression scale.  To ensure that $r_s$ information is not informing the $H_0$ constraints, one could employ strategies such as marginalizing over the locations of the BAO and baryon suppression features (similar to what we have done in the Fisher analysis presented above), although more development is required before such a procedure can be applied to a high-precision data  analysis.  We will defer the discussion and application of such methods to future work.

\section{Summary}
\label{sec:discussion}

We have shown how measurements of CMB lensing can be combined with external data to obtain constraints on the Hubble constant that --- unlike constraints derived from the primary CMB or from BAO --- do not depend on the sound horizon scale at CMB last-scattering $r_s$.  Fundamentally, the information about $H_0$ derived here comes from the dependence of the CMB lensing power spectrum on the (projected) scale of the horizon size (or Jeans scale) at matter radiation equality. Using several tests and arguments we have demonstrated that, since projection washes out baryonic features, such CMB lensing power spectrum constraints do not rely on $r_s$.

With current measurements of the CMB lensing power spectrum from \citet{Plancklensing2015}, a CMB-motivated (but conservative and $r_s$-independent) prior on $A_s$, and a supernova-based constraint on $\Omega_m$, the Hubble constant is constrained to $H_0 = 73.5\pm 5.3\,{\rm km}/{\rm s}/{\rm Mpc}$.  This value is consistent with the CDL-inferred value at roughly $0.1\sigma$ and with the CMB-inferred value at $1.2\sigma$ (where $\sigma$ here represents the quadrature sum of the errors from both measurements). While obtaining a best fit Hubble constant that is somewhat higher than {\it Planck} and agrees with the local CDL measurements is intriguing, the errors are still large enough that our results are consistent with both probes; an interpretation of this result as evidence of new physics would, of course, be very premature. Nevertheless, this result certainly motivates further efforts to improve the constraints from our method.

With future data, we find that an $r_s$-independent constraint on $H_0$ at the level of 3~km/s/Mpc
can be obtained by combining CMB lensing measurements with priors on $\Omega_m$ (from either supernovae or galaxy lensing) and $A_s$ (from the CMB power spectrum).  This is level of uncertainty is certainly interesting, but is not tight enough to definitively resolve the Hubble tension found in current datasets.  The $H_0$ constraints could be improved further with higher sensitivity measurements of the lensing power spectrum, as long as the constraints on $\Omega_m$ and either $A_s$ or $n_s$ are tightened as well. 

However, it is apparent that the $H_0$ constraints from this method only improve fairly slowly as CMB surveys become more powerful. Therefore, perhaps the most promising avenue to improve our analysis of current data is to obtain a similar $k_{\rm eq}$-derived Hubble constant measurement from the 3D galaxy power spectrum broadband shape. Such an analysis poses considerable challenges, as it is currently not clear how to rigorously remove the sound horizon information which enters through baryonic effects. We defer a detailed treatment of such an analysis to future work.

\section*{Acknowledgements}

We thank Graeme Addison, Colin Hill, Lloyd Knox, Oliver Philcox, and Marko Simonovi\'c for comments on our draft.  We thank Vivian Miranda, Anthony Challinor and Antony Lewis for useful discussions.

\section*{Data availability}

The data used to generate the figures in this work are available upon request.

\bibliographystyle{mnras}
\bibliography{thebib}

\vspace{-0.2cm}

\appendix

\section{Accuracy of lensing approximations and physical intuition}
\label{app:lensing_approx}

\begin{figure}
    \centering
    \includegraphics[scale=0.4]{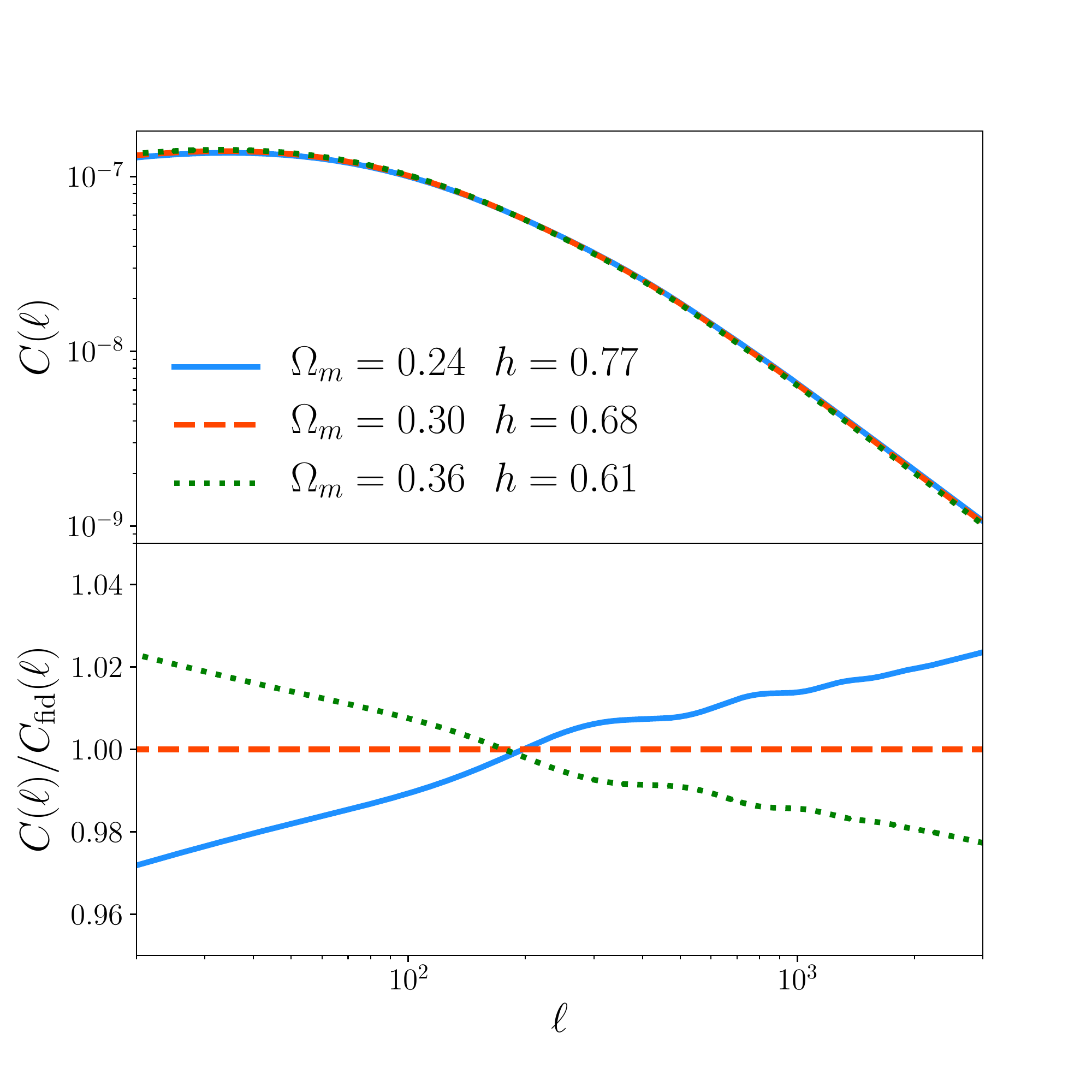}
    \caption{The CMB lensing power spectrum for different values of $\Omega_m$ and $h$, with the values chosen so as to preserve the parameter combination $L_{\rm eq} \propto \Omega_m^{0.6} h$.  The bottom panel shows the lensing power spectra relative to that computed for $\Omega_m = 0.3$ and $h= 0.68$.  The dependence of the CMB lensing power spectrum on $\Omega_m$ and $H_0$ is largely captured by its dependence on $L_{\rm eq}$.  Preserving $L_{\rm eq}$ keeps the lensing power spectrum constant at roughly 3\%.  We have included the impact of baryons, nonlinear evolution, and neutrinos on the matter power spectrum when generating this figure.}
    \label{fig:cmblensing_test}
\end{figure}

In \S\ref{sec:parameter_dependence} we argued that the dependence of the CMB lensing power spectrum on cosmological parameters could be largely captured via its dependence on $L_{\rm eq}$ (and $A_s$).  In  Fig.~\ref{fig:cmblensing_test} we test this result by directly computing the CMB lensing power spectrum for different $\Omega_m$, scaling $H_0$ in order to preserve $L_{\rm eq}$.  For the remaining cosmological parameters, we adopt the fiducial values from Table~\ref{tab:priors}; we include the impact of nonlinear effects on the matter power spectrum via \texttt{HALOFIT} \citep{Smith:2003}, and also assume a minimal neutrino mass, included via the \citet{Bird:2012} extension to \texttt{HALOFIT}.  We find even with the effects of baryons, nonlinear evolution and neutrinos taken into account, preserving $L_{\rm eq}$ when varying $\Omega_m$ is sufficient to recover the original power spectrum to better than 4\% over a wide range of $L$.  

We can gain physical intuition for the $A_s$ and $L_{\rm eq}$ dependence of the CMB lensing power spectrum as follows (see also the discussion Appendix E of \citealt{Plancklensing2015}).  For a given scale, $k$, the number of lenses of that scale between the observer and the source plane is $k \chi_*$.  Increasing $k_{\rm eq}$ will shift the power spectrum to higher $k$, causing lenses of a given potential depth to be at smaller scales.  To preserve the number of lenses of each depth to the source while increasing $k_{\rm eq}$, one should therefore decrease $\chi_*$ proportionately.  Consequently, we can preserve the amplitude of the lensing power by fixing $k_{\rm eq} \chi_* \equiv L_{\rm eq}$.  We can also understand the dependence of the shape of the power spectrum on $L_{\rm eq}$.  Since the lensing power spectrum receives its largest contribution from around $\chi_*/2$, the observed angular size of a typical deflection will be inversely proportional to  $\chi_*$.  If we increase  $k_{\rm eq}$, we make lenses of a given potential depth smaller by shifting the entire power spectrum to higher $k$.  Thus, fixing $k_{\rm eq}\chi_* \equiv L_{\rm eq}$ will also roughly preserve the angular scale lenses of a given potential depth, thereby preserving the shape of the observed lensing power spectrum. 

The $A_s$ dependence of the CMB lensing power spectrum can be understood trivially from the fact that $A_s$ scales the amplitude of the linear power spectrum, $P_{\rm lin}(k)$, and $P(k)$ enters linearly into the Limber approximation.  More physically, increasing $A_s$ by some factor $f$ will increase the amplitude of the overdensity fluctuations on a given scale by a factor of $\sqrt{f}$.   Since the photon deflections scale with the overdensity, they will also get larger by $\sqrt{f}$, causing the deflection power to increase by $f$.  Consequently, the CMB lensing power will scale linearly with  $A_s$.  

\bsp	
\label{lastpage}
\end{document}